\documentclass[10pt]{iopart}

\usepackage{subcaption}
\usepackage{array}
\usepackage{multirow}
\usepackage{booktabs}
\usepackage{textcomp}
\usepackage{miller}
\usepackage{xfrac}
\usepackage{mathrsfs}
\usepackage{ulem}
\PassOptionsToPackage{hyphens}{url}
\usepackage[colorlinks=true]{hyperref}
\usepackage{orcidlink}
\newcommand{\textapprox}{\raisebox{0.5ex}{\texttildelow}}

\usepackage{hypernat}

\begin{document}
\frenchspacing

\title{Biased self-diffusion on Cu surface due to electric field gradients}

\author{Jyri Kimari$^1$, Ye Wang$^2$, Andreas Kyritsakis$^2$, Veronika Zadin$^2$ and Flyura Djurabekova$^1$}

\address{$^1$ Helsinki Institute of Physics and Department of Physics, P.O. Box 43 (Pietari Kalmin katu 2), FI-00014 University of Helsinki, Finland}
\address{$^2$ Institute of Technology, University of Tartu, Nooruse 1, 50411 Tartu, Estonia}
\ead{jyri.kimari@helsinki.fi}

\begin{abstract}

  Under strong electric fields, an arc of strong current flowing through plasma can link two metal surfaces even in ultra high vacuum. Despite decades of research, the chain of events leading to vacuum arc breakdowns is hitherto unknown. Previously we showed that a tall and sharp Cu nanotip exposed to strong electric fields heats up by field emission currents and eventually melts, evaporating neutral atoms that can contribute to plasma buildup.
  
  In this work, we investigate by means of molecular dynamics simulations whether surface diffusion biased by the presence of an electric field gradient can provide sufficient mass transport of atoms toward the top of the nanotip to maintain supply of neutrals for feeding plasma. To reach the necessary timescales and to add electric field in MD, we utilized a novel combination of collective variable~-driven hyperdynamics acceleration and coupling to a finite element mesh. In our simulations, we observed biased self-diffusion on Cu surfaces, that can contribute to the continuous replenishment of particle-emitting nanotips. This mechanism implies a need to reduce the rate of surface diffusion in devices that are susceptible to vacuum arcs. Finding suitable alloys or surface treatments that hinder the observed biased diffusion could guide the design of future devices, and greatly improve their efficiency.

\end{abstract}


\submitto{\JPD}

\maketitle

\ioptwocol

  \section{Introduction}
  \label{sec:intro}
    
    Metallic surfaces become subjected to unprecedentedly high electric fields in the high-power devices of improved efficiency but compact dimensions. Vacuum is known for very high insulating properties. The higher the vacuum the higher electric fields can be applied between the two metal plates before an arc will bridge them. Hence there is always a certain threshold voltage (known as the breakdown voltage), at which a medium conducting strong arcing currents appears even in ultra high vacuum. The Compact Linear Collider~(CLIC)~\cite{clic2016updated}, a proposed next-generation particle accelerator in CERN, is one of the important examples where tiny vacuum arcs may affect the performance efficiency of the entire machine. This room-temperature Cu ``tube'' spanning from a few millimeters in inner diameter to~50 kilometers in length is designed to enable up to~3\,TeV collision between electrons and positrons. Both types of particles are accelerated to the required energies by high-gradient electromagnetic fields within Cu accelerating structures. The bursts of vacuum arcs consume power and divert bunches of accelerated particles, as well as damage the accelerating structures themselves. Desirable reduction in the occurrence of these current bursts is difficult, since their mechanisms are not completely known.
    
    Under high electric field, some electrons always leak into the vacuum through the field emission process. These initially low currents rise by many orders of magnitude, when a plasma builds up above the surface~\cite{zhou2020spectroscopic}. To form plasma, particles of both negative (electrons) and positive (ions) charges are needed. The positive ions are thought to originate from the surfaces exposed to the electric field. The electric field magnitudes applied in vacuum arcing experiments---in the hundreds of~MV/m~\cite{saressalo2020classification}---are too low for direct field evaporation, which takes place in the~10--50\,GV/m range~\cite{kelly2012atom}. Hence it has been suggested that a feedback-loop of self-reinforcing growth of a surface protrusion must exist~\cite{pohjonen2011dislocation, kyritsakis2018thermal}. The sharper the tip, the stronger the field at its top. The enhanced field induces stronger currents eventually leading to melting and subsequent evaporation of neutral atoms and atom clusters into vacuum. However, the protrusion growth is expected to be too fast for experimental observation, hence, theoretical and computational models are developed to understand the mechanisms governing the process of surface protrusion growth.
    
    Microscopy of surfaces that have experienced multiple breakdowns reveal a large number of breakdown spots in the shape of solidified molten regions known as craters, see e.g. Refs.~\cite{saressalo2020classification,shipman2015experimental,saressalo2021situ}. The crater edges are jagged, and thus they themselves can function as field enhancing features~\cite{saressalo2020classification}. However, the field enhancement on such features is weak as the features are generally blunt---the aspect ratio of the crater edge features has not been reported to be sufficiently high. Hence, these frozen-in features cannot initiate a feedback loop that can result in a subsequent breakdown. Thus, some additional mechanisms of growth and sharpening of protrusions must exist.
    
    Under an applied electric field $F_\mathrm{ext}$, any surface asperity induces local field enhancement that is estimated as $F=\beta F_\mathrm{ext}$, where $\beta \approx h/r$, the geometric aspect ratio of the surface asperity~\cite{edgcombe2001enhancement,edgcombe2001microscopy,djurabekova2011atomistic}. Naturally, the enhanced field may result in enhanced Maxwell tensile stress, which will affect locally the atomic dynamics at this enhancing surface feature~\cite{parviainen2015atomistic}. However, this is not the only effect, which can be caused by the applied electric field at the surface. Already in~1975, Tsong et al. proposed a mechanism where atomic diffusion is biased toward stronger electric fields in the presence of an electric field gradient~\cite{tsong1975direct}. This bias is expected due to alternation of polarization characteristics of surface atoms, such as dipole moments and polarizability. Recently we have improved this approach by applying the theory to the changes in the dipole moment and the polarizability of the entire surface due to a single jump of a migrating adatom~\cite{kyritsakis2019atomistic}. The bias can be expressed as a change in the migration energy barrier $E_\mathrm{m}$:
    \begin{equation}
      \label{eq:modbarrier}
      \Delta E_\mathrm{m} = -\mathcal{M}_\mathrm{sl} F - \frac{\mathcal{A}_\mathrm{sl}}{2}F^2 - \mathcal{M}_\mathrm{sr}\gamma l - \mathcal{A}_\mathrm{sr}\gamma l F
    \end{equation}
    $F$ is the strength of the the electric field at the initial lattice site of the atom, and $l$ is the distance between the initial site and the saddle point, i.e. the highest energy point along the minimum energy path of the jump. $\mathcal{M}$ is the dipole moment and $\mathcal{A}$ is the polarizability of the system; these electrical parameters are material-dependent. Subscript sl denotes a difference between the lattice site and the saddle point (e.g. $\mathcal{M}_\mathrm{sl}\equiv\mathcal{M}_\mathrm{s}-\mathcal{M}_\mathrm{l}$), and sr the difference between the saddle point and the reference system of a flat substrate without the adatom. Note that while the electric field affects the material in the direction of surface normal (along the direction of local field), the electric field gradient will affect the surface in the direction perpendicular to the surface normal and, hence, to the electric field. The barrier is generally lower moving in the direction of the gradient, and higher moving in the opposite direction.

    By the mechanism of biased diffusion under electric field gradient, a field enhancing feature (a tip for short) would tend to grow taller and sharper, thus also producing a higher factor $\beta$. This phenomenon has been observed in kinetic Monte Carlo~(KMC) simulations of W surface by Jansson et al.~\cite{jansson2020growth}.

    Although a KMC approach to model surface diffusion processes is very attractive, we have previously identified numerous challenges in KMC simulations of the Cu surface~\cite{baibuz2018migration,kimari2020application}. A significant part of these issues are caused by the assumption of a rigid lattice with fixed lattice sites; on the face-centered cubic~\hkl{111} surface, for instance, the off-lattice hexagonal close-packed~(hcp) sites often have a stability comparable with a regular lattice site.  Nevertheless, the understanding of the breakdown phenomenon requires simulation results under these conditions.
    
    In this work, we study the drift of Cu adatoms under inhomogeneous electric field with the electric field gradient due to existing surface features. We carry out these simulations using molecular dynamics~(MD), since it offers the flexibility of dynamically evolving system with all positions accessible within the simulation cell. To introduce the effect of electric fields and to overcome length and time scale limitations, we modified the classical MD by coupling it to finite elements method~(FEM)~\cite{veske2018dynamic} field solver, and applying collective variable~-driven hyperdynamics~(CVHD)~\cite{bal2015merging} acceleration. To verify the validity of the~MD-FEM electrostatic model, we estimate the dipole moment and polarizability characteristics based on the diffusion results, and compare them to the corresponding values calculated with density functional theory~(DFT). All simulation details are described in Sec.~\ref{sec:methods}. Sec.~\ref{sec:results} describes the simulation results, which are further discussed in  Sec.~\ref{sec:discussions}. Finally, conclusions are drawn in Sec.~\ref{sec:conclusions}.
    
  \section{Methods}
  \label{sec:methods}

    \subsection{Molecular dynamics}
    \label{subsec:methods}

\begin{figure*}
        \centering
        \hfill
        \begin{subfigure}{0.2\linewidth}
          \includegraphics[width=\linewidth,trim={2cm 0 2cm 0}, clip]{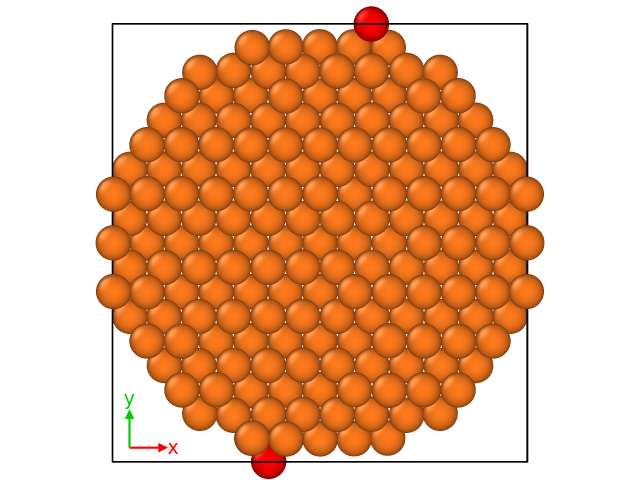}
        \end{subfigure}
        \hfill
        \begin{subfigure}{0.21\linewidth}
          \includegraphics[width=\linewidth,trim={2cm 0 2cm 0}, clip]{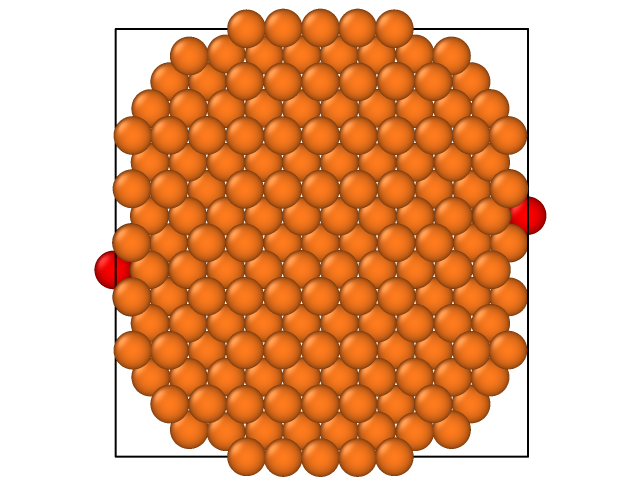}
        \end{subfigure}
        \hfill
        \begin{subfigure}{0.28\linewidth}
          \includegraphics[width=\linewidth]{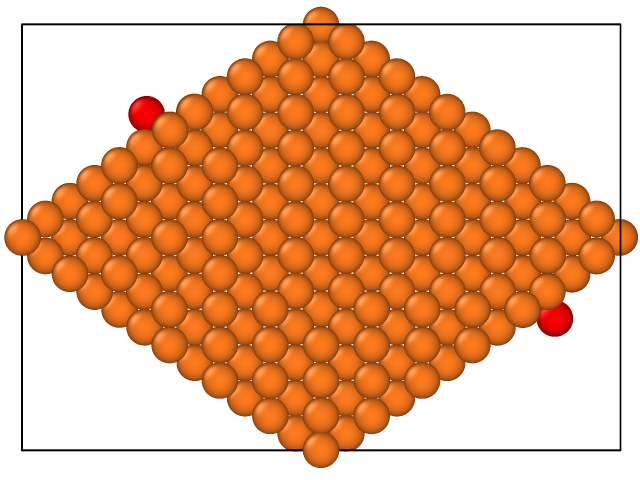}
        \end{subfigure}
        \hfill{}
        \caption{Cross-sections of the MD-simulated nanowires of this work. The added adatoms are colored red, with the rest of the atoms copper-colored. Adatoms were added either on the~\hkl{100} (left), the~\hkl{110} (middle), or the~\hkl{111} (right) surfaces of the wire.}
        \label{fig:sections}
      \end{figure*}

      We simulated Cu self-diffusion along a nanowire surface with molecular dynamics~(MD), using the LAMMPS software~\cite{plimpton1995fast}. The MD region consisted of a slice of nanowire with periodic boundary conditions along the length of the wire, which was set in the~\hkl<110> direction, in terms of Miller indices. The thickness of the nanowire slice was~8 interatomic distances, i.e.~\textapprox 20\,\AA, and the radius of the wire ranged from~10 to~20\,\AA. Two adatoms were added on the surface of the wire, to be tracked for the total distance they travel along the wire length during the simulation. We studied diffusion on the three lowest-index surfaces: the~\hkl{100}, the~\hkl{110}, and the~\hkl{111} surface. The cross-section of the wire was roughly circular in the~\hkl{100} case, but modified to increase the area of the facet of interest in the~\hkl{110} and the~\hkl{111} cases to reduce the probability of atoms leaving this facet, which might affect the statistical analysis. The geometries with different cross-sections of the wires are shown in Fig.~\ref{fig:sections}. Ten repetitions of each case were conducted with different random seeds and randomized positions of the adatoms on the desired surfaces.

      The atomic interactions were defined by an MD/MC-CEM potential by Stave et al.~\cite{stave1990corrected}. This potential was optimized to reproduce the properties of Cu surfaces~\cite{sinnott1991corrected}, such as the surface energies and relaxation effects. We used a~4\,fs timestep, and a Nosé-Hoover thermostat set to~300\,K. Simulation systems were thermalized for~400\,ps before any other modifications. The atoms immediately around the axis of the nanowire were fixed to prevent any overall drift due to external forces. Input files that define the MD region geometry and all other simulation details are included in supplementary material.

      The total runtimes of the simulations were~10--700\,\textmu s when collective variable~-driven hyperdynamics~(CVHD; see Sec.~\ref{subsec:cvhd}) was used, and~10--100\,ns otherwise.

      Beyond the~20\,\AA\ thick MD region, the simulated system was extended with a continuum surface mesh, to allow a more realistic calculation of the electric field with the finite elements method~(FEM), described in the section \ref{subsec:fem}.

    \subsection{Finite elements method}
    \label{subsec:fem}

      We obtained the distribution of the electric field in our simulations by using the Femocs~\cite{veske2018dynamic} library that is a finite element method~(FEM) solver. The library has interfaces to LAMMPS and Parcas~\cite{nordlund1997point} MD softwares and the Kimocs~\cite{jansson2016long} kinetic Monte Carlo~(KMC) software, and it can also be compiled as a standalone program. Femocs translates the applied electric field into a potential that follows the metallic equipotential surfaces, and calculates the surface charges. Charge is distributed to surface atoms that consequently experience electrostatic forces added within the MD algorithm.

      Femocs can easily extend the solver mesh beyond the atomic system, which increases significantly the simulation domain compared to limited length scales of MD. To emulate the situation of surface diffusion under electric field \emph{gradient}, we built the Cu system with a nanotip placed on a surface. The nanotip is sufficiently tall to enhance the electric field toward its sharp top, imposing a ``natural'' gradient along its length. Specifically, we are interested in the $rz$-component of the gradient tensor, i.e. the partial derivative of the electric field radial component with respect to the (axial) $z$-coordinate:
      \begin{equation}
        \label{eq:gammatensor}
        \gamma_{rz} = \frac{\partial F_r}{\partial z},
      \end{equation}
      
\begin{figure}
        \centering
        \includegraphics[width=0.9\linewidth]{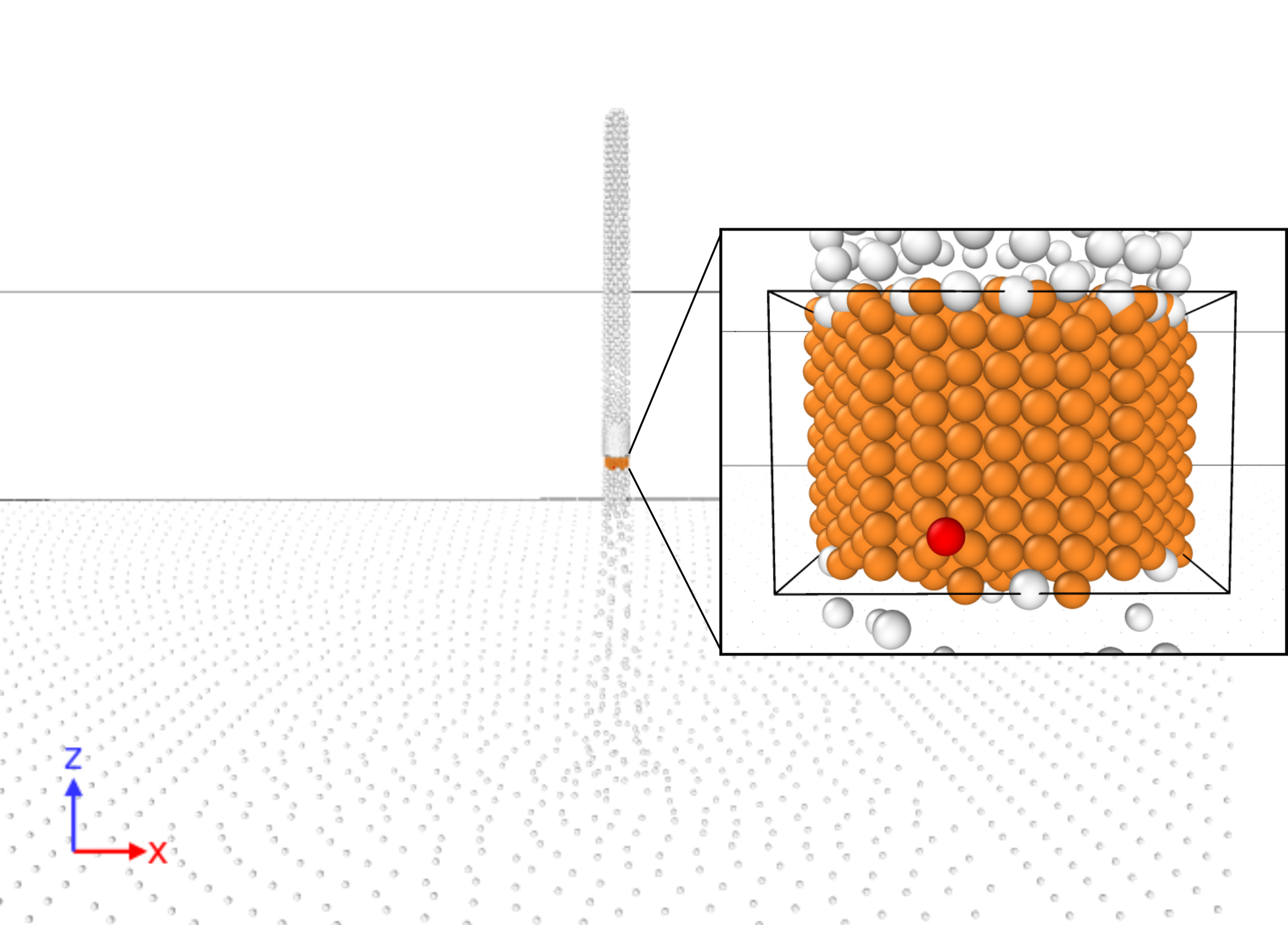}
        \caption{Extended simulation system, with the MD region (zoomed into in the inset) in copper color and the static, continuous extension in gray. The system has periodic boundaries in the horizontal directions, and open boundaries in the $z$-direction. The MD region, on the other hand, has a periodic boundary through its own $z$-span, independently of the extension.}
        \label{fig:extension}
      \end{figure}

      Two different nanotip geometries were used: a tall,~93\,nm tip, and a short,~5\,nm one, with a similar cross-section shape as the MD region in each case. In the case of nearly elliptical~\hkl{100} nanowire, we used elliptical extension cross-section; for the more flattened~\hkl{110} wire, we applied similar flattening to the extension cross-section; and for the~\hkl{111} wire, we used a diamond-shaped cross-section for the extension. The MD region was placed either in the bottom, the middle or the top of the tall nanotip, or in the middle of the short one. See Fig.~\ref{fig:extension} for an example of the extended tall nanotip system. While the MD region has periodic boundaries in the $z$-direction and open boundaries in the horizontal directions, \emph{the extended simulation box has open boundaries in the $z$-direction, and periodic boundaries in the horizontal directions}.

      We want to frame the simulation setting in the following way. The property of interest is the drift of single adatoms along the $z$-coordinate. To collect statistics, one option would be to add adatoms e.g. in the middle of the wire, and remove any atoms that exit through the top or the bottom of the MD region. What we have done here, instead, is a more convenient way to accomplish this by utilizing periodic boundaries. Atoms crossing the MD boundary can be, for all intents and purposes, considered to be removed from the system. At the same instant, a new adatom is added in the system from the opposing boundary---the fact that this is technically the same adatom does not make a difference. We are \emph{not} simulating the diffusion of atoms along a \emph{long} extended region, but rather investigating the \emph{local bias} that is imposed on the adatoms by the electric field gradient in the thin MD slice.
      
      We want to emphasize that when an atom is arbitrarily close to the top (bottom) of the MD region, it does not see any fields present at the bottom (top) of the region. Instead, it sees the fields present in the extended system above (below) the MD region. Once the crossing of the border happens, the atom lands in a new environment, forgetting the fields it just left. Even if the migration behavior were anomalous precisely at the border of the MD region, this would constitute a small error in the total drift through numerous MD region heights. Furthermore, regardless of whether the possible boundary anomaly was attractive (adatoms are less likely to move away from the boundary) or repulsive (adatoms are less likely to cross the boundary), the error is expected to be in the \emph{downward} direction (reducing the drift). Thus, any error in the boundary will not result in a false positive result for the observation of biased diffusion.
      
      To solve the electric field, Femocs first constructs a surface mesh based on the atomic coordinates and the continuous extension. It also constructs a 3D mesh for the vacuum. Neglecting any space charge that would be due to electron emission or Cu ions detached from the surface, the Laplace equation holds:
      \begin{equation}
        \label{eq:laplace}
        \nabla^2 \Phi = 0
      \end{equation}
      $\nabla^2$ is the Laplace operator that gives the divergence of the electric field that is due to the electrostatic potential $\Phi$. In the absence of any total charge in the system, the divergence has to equal zero everywhere.

      The boundary conditions~(BC) used for the Laplace equation are in this case:
      \begin{enumerate}
        \item Constant $\Phi$ everywhere on the surface, since the Cu is a conductor (Dirichlet BC).
        \item Constant $\nabla \Phi$ at the top of the extended simulation box due to a far-away anode (Neumann BC).
        \item No electric flux through the extended simulation box boundaries.
      \end{enumerate}
      All these boundary conditions are implemented in Femocs. It solves the Laplace equation with FEM in the 3D mesh generated in the vacuum, bounded by the surface.

      Solving the potential $\Phi$ lets us distribute the surface charges to individual surface atoms (see Ref.~\cite{veske2018dynamic} for details) and calculates the electric field
      \begin{equation}
        \label{eq:field}
        \mathbf{F} = -\nabla \Phi
      \end{equation}
      The electric field exerts forces to the charged atoms. The forces are finally exported back to the MD algorithm to modify the atomic dynamics accordingly.

      To make the simulation more efficient, the FEM mesh (and thus the electric field) is only updated when the root mean square~(RMS) displacement of surface atoms is sufficiently large compared to the previous mesh generation. We used a maximum RMS value of 0.38\,\AA, found to provide a good trade-off between accuracy and efficiency~\cite{veske2018dynamic}. The atomic partial charges and electrostatic forces are updated at every timestep.

      The applied electric field range studied in this work starts from~100\,MV/m. The upper limit of the field depends on the geometry: due to the field enhancement near the tip of the~93\,nm tall protrusion, the atomic structure disintegrates at applied electric fields above~1\,GV/m. At the bottom of the protrusion, as well as in the~5\,nm protrusion, the field could be increased up to~5\,GV/m. For reference, the nominal acceleration voltage in the CLIC device is~100\,MV/m, corresponding to fields of~200\,MV/m or more on the surfaces surrounding the beam~\cite{saressalo2020classification}.
      
      The direction of the applied field was exactly opposite to the $z$-axis, so that the simulation model acted as the cathode of a two-electrode electrostatic system. However, we note that we neglect the processes of electronic heating and space charge effects at the cathode to be able to focus on biased diffusion due to an electric field; hence, the choice of the field direction is not critical for the current simulations.

      Input files detailing the FEM part of the simulations are available in supplementary material. This includes both the FEM solver parameters and the FEM extension geometries for all simulated cases.
  
    \subsection{Collective variable -driven hyperdynamics}
    \label{subsec:cvhd}

      Diffusion process timescales are often beyond the range accessible by MD. In the scope of this work, the timescale problem applies to the Cu~\hkl{100} surface. On the~\hkl{110} and the~\hkl{111} surfaces the potential energy surface felt by adatoms is smooth enough for fairly easy transitions between the lattice sites. However, on the~\hkl{100} surface the adatoms sit in deep potential energy wells, with approx.~0.5\,eV activation energy for migration.
      
      To assess the effect of biased diffusion on all three most commonly appearing surfaces, we use the collective variable -driven hyperdynamics~(CVHD) acceleration~\cite{bal2015merging} for the surface with the ~\hkl{100} orientation. Since the details of the algorithm vary slightly between different implementations~\cite{fukuhara2020accelerated}, we briefly review the basics of this acceleration method and its parameters for better reproducibility of the presently reported results.
      
      In CVHD, the potential energy of the system is biased by a term that depends on a one-dimensional collective variable~(CV) $\eta$:
      \begin{equation}
        \label{eq:Vbias}
        V^*(\mathbf{R}) = V(\mathbf{R}) + \Delta V(\eta)
      \end{equation}
      The term $V(\mathbf{R})$ is the regular interatomic potential in the system, a function of the atomic coordinates $\mathbf{R}$, and $\Delta V(\eta)$ is the added CVHD bias. The $\eta$ variable is chosen such that it is able to detect the rare events that are of interest, i.e. $\eta$ should have almost zero values when the system as a whole is near equilibrium, and values near unity when the system is almost at the boundary between the two states separated by the event; $\Delta V(\eta)$ is an almost monotone decreasing function of $\eta$, pushing the system away from the equilibrium. Presently by the events we understand migration jumps of individual adatoms between the neighboring lattice sites. We adopt the ``bond-breaking'' based CV following the suggestion by  Bal and Neyts~\cite{bal2015merging}. Starting from the interatomic distances between all nearest neighbors~(NN) in the system, $r_i$, the CV is defined in the following way. First, each $r_i$ is associated with a \emph{local distortion}:
      \begin{equation}
        \label{local_distortion}
        \chi_i =
        \left\{\begin{array}{@{}rl@{}}
          0                                                        , &\text{if}\quad r_i \leq r_\mathrm{min}\\
          \frac{r_i-r_\mathrm{min}}{r_\mathrm{max}-r_\mathrm{min}} , &\text{if}\quad r_\mathrm{min} < r_i \leq r_\mathrm{max}\\
          1                                                               , &\text{if}\quad r_i > r_\mathrm{max}
        \end{array}\right.
      \end{equation}
      Here, $r_\mathrm{min}$ and $r_\mathrm{max}$ are user-defined parameters, bounding the interval within which $r_i$ is expected to stretch due to thermal motion. $r_\mathrm{min}$ is usually set to the equilibrium distance between the nearest neighbors at the given temperature of the simulated domain, and $r_\mathrm{max}$ is set to be the "breaking bond" distance, where one of the atoms in the pair $i$ has jumped away from its original lattice site.
      
      The next step toward a \emph{collective} variable is to define the global distortion as a $p$-norm of the local distortions:
      \begin{equation}
        \label{eq:chi}
        \chi = \left( \sum_i \chi_i^p \right)^\frac{1}{p}
      \end{equation}
      $p>1$ is another user-defined parameter that is designed to emphasize the effect of large individual local distortions $\chi_i$ within the global distortion, i.e. to be more sensitive to individual processes anywhere in the system. The higher the value of $p$, the more sensitive the method is to the significant distortions.
      
      Finally, to bound the collective variable to interval $[0,\,1]$, the global distortion is passed through the cosine function:
      \begin{equation}
        \label{eq:eta}
        \eta = 
        \left\{\begin{array}{@{}rl@{}}
          \frac{1}{2}\left[1 - \cos(\pi \chi^2) \right], & \text{if}\quad\chi \leq 1\\
          1                                            , & \text{if}\quad\chi > 1
        \end{array}\right.
      \end{equation}
      This is the CV that appears in the bias potential $\Delta V(\eta)$. The nature of this CV is such that it will have small values when \emph{all} the NN distances $r_i$ in the system are close to $r_\mathrm{min}$, and large values when at least one $r_i$ is stretched. When any $r_i > r_\mathrm{max}$, the CV will cap to exactly~1. By monitoring this capping, transitions in the system can be detected and the NNs re-assigned when necessary.

\begin{figure}
        \centering
        \input{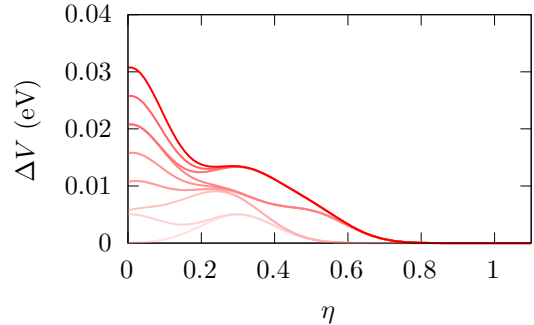}
        \caption{Schematic of the evolution of the bias potential over $10\tau$ steps. Small Gaussian hills are summed together to form the total bias potential function (the topmost line, in solid red). The bias pushes the system toward higher values of $\eta$, i.e. away from the current state.}
        \label{fig:DeltaV}
      \end{figure}

      The bias potential is constructed dynamically during the simulation. In the beginning, $\Delta V(\eta) = 0$ everywhere. At user-defined intervals $\tau$, the current value of $\eta=\eta_\tau$ is calculated. At this location, \emph{a small Gaussian hill} is added, and $\Delta V(\eta)$ will become
      \begin{equation}
        \label{eq:DeltaV}
        \Delta V(\eta) = w \exp\left[ -\frac{(\eta - \eta_\tau))^2}{2\delta^2} \right]
      \end{equation}
      After this, at every step $k\tau$, ($k=1,\,2,\,3,\,\ldots$), a similar Gaussian hill is added at the location $\eta_{k\tau}$. Parameters $w$ and $\delta$ are defined by the user and they control the height and the width of the potential energy ``packages'' that are added to $\Delta V$. This way, the bias potential grows slowly over time as illustrated in Fig.~\ref{fig:DeltaV}.
      
      In the well-tempered metadynamics~\cite{barducci2008well} variant of CVHD, used also in this work, the height of the Gaussian hills is modified so that lower
      energies are added in $\eta$ regions where the bias is already high:
      \begin{equation}
        \label{eq:temper}
        w_k = w \exp\left(-\frac{\Delta V(\eta_{k\tau})}{k_\mathrm{B} \Delta T}\right)
      \end{equation}
      where $k_\mathrm{B}$ is the Boltzmann constant and $\Delta T$ is an algorithmic bias temperature---it has no connection to the physical temperature of the system.

      The final equation required for the CVHD acceleration is the stretching of time due to the bias potential. Every timestep length $\Delta t_\mathrm{MD}$ (that is used in the integration of the equations of motion), is multiplied by a bias-dependent factor to account for the time that would elapse if the transition had happened naturally due to thermal vibrations:
      \begin{equation}
        \label{eq:CVHD_time}
        \Delta t_\mathrm{CVHD} = \Delta t_\mathrm{MD} \left\langle \exp\left(\frac{\Delta V(\eta)}{k_\mathrm{B}T}\right) \right\rangle
      \end{equation}
      where $T$ is the temperature of the simulated system. In other words, the more bias has been accumulated, the faster the time advances. This somewhat resembles KMC dynamics, where the system outright skips any movement between interesting jumps.
      
      In the beginning of the simulation, all NN distances are close to the equilibrium value $r_\mathrm{min}$, and thus $\eta$ will have values near to zero. This will cause bias $\Delta V(\eta)$ grow specifically at low values of $\eta$. The bias potential will exert a force that drives $\eta$ to higher values, i.e. where at least one NN distance---``bond''---is stretched. Due to the exponent $p$ in Eq.~\eref{eq:chi}, small distortions will have little effect in $\eta$, and thus the bias will not significantly affect atoms that are tightly bound in the lattice sites, i.e. the lengths of their bonds do not deviate  from equilibrium significantly due to thermal motion. Atoms on the surface, on the other hand, have more space to move, and feel the bond-stretching force of the bias.

      Over time, $\eta$ will be sampled at higher and higher values, making the bias push stretching bonds even further, until one atom in the system finally makes a jump further away from its NN than the distance $r_\mathrm{max}$. This causes $\eta$ to saturate at value~1 ``indefinitely''. If $\eta=1$ for $\tau_\mathrm{threshold}$ timesteps, the system is considered to have moved to a new state. At this point,
      \begin{enumerate}
        \item The NN atoms will be recalculated. Any atoms that have distance less than $r_\mathrm{cut}$ are considered NNs with each other.
        \item All bias potential is removed, i.e. the accumulation of $\Delta V$ begins anew.
      \end{enumerate}
      This ensures that no bias is added in the transition states of the system, which  would skew the dynamics. The bias resetting allows CVHD avoiding the problem of small barriers, that many other acceleration methods face~\cite{bal2015merging}. Even if the dynamics of the system are unknown beforehand, with widely varying migration energy barriers, all processes that have a barrier higher than the Gaussian height $w$ of Eq.~\eref{eq:DeltaV} will be handled correctly.
      
      The resetting of the bias may decrease the acceleration efficiency in simulations where events happen very frequently. This is another reason why we did not use CVHD in the~\hkl{110} and~\hkl{111} surface simulations, where diffusion is very fast---the obtained boost was less than the additional cost of the CVHD algorithm. The~\hkl{100} surface, on the other hand, is truly ideal for adjusting the expected frequency of jumps to utilize dynamic CVHD to its full potential. Extending the MD region with the FEM mesh allows us to further decrease the number of atoms (\textapprox frequency of jumps) in the system without encountering detrimental finite size effects.

      The LAMMPS software includes the CV framework as a standard feature. The implementation of the bond-breaking CV and the CVHD acceleration (bias potential and the time factor) was written by Bal and Neyts~\cite{bal2015merging} and updated for a newer version of LAMMPS by Kurki~\cite{kurki2020performance}. The parameters used in this work are tabulated in Tab.~\ref{tab:CVHD}. Input files defining the CVHD behavior are also provided in the supplementary material.
      \begin{table}
        \centering
        \caption{Parameters used in the CVHD acceleration.}
        \label{tab:CVHD}
        \begin{tabular}{lc}
          \toprule
          Parameter                 & Value        \\
          \midrule
          $r_\mathrm{min}$          & 2.56466\,\AA \\
          $r_\mathrm{max}$          & 3.30\,\AA    \\
          $r_\mathrm{cut}$          & 3.00\,\AA    \\
          $p$                       & 20           \\
          $w$                       & 0.005\,eV    \\
          $\delta$                  & 0.05         \\
          $\tau$                    & 4\,ps        \\
          $\tau_\mathrm{threshold}$ & 10\,ps       \\
          $\Delta T$                & 2000\,K      \\
          \bottomrule
        \end{tabular}
      \end{table}
      
    \subsection{Calculation of surface polarization characteristics by density functional theory}
    \label{DFT}
    
    \begin{figure}
        \centering
        \includegraphics[width=\linewidth]{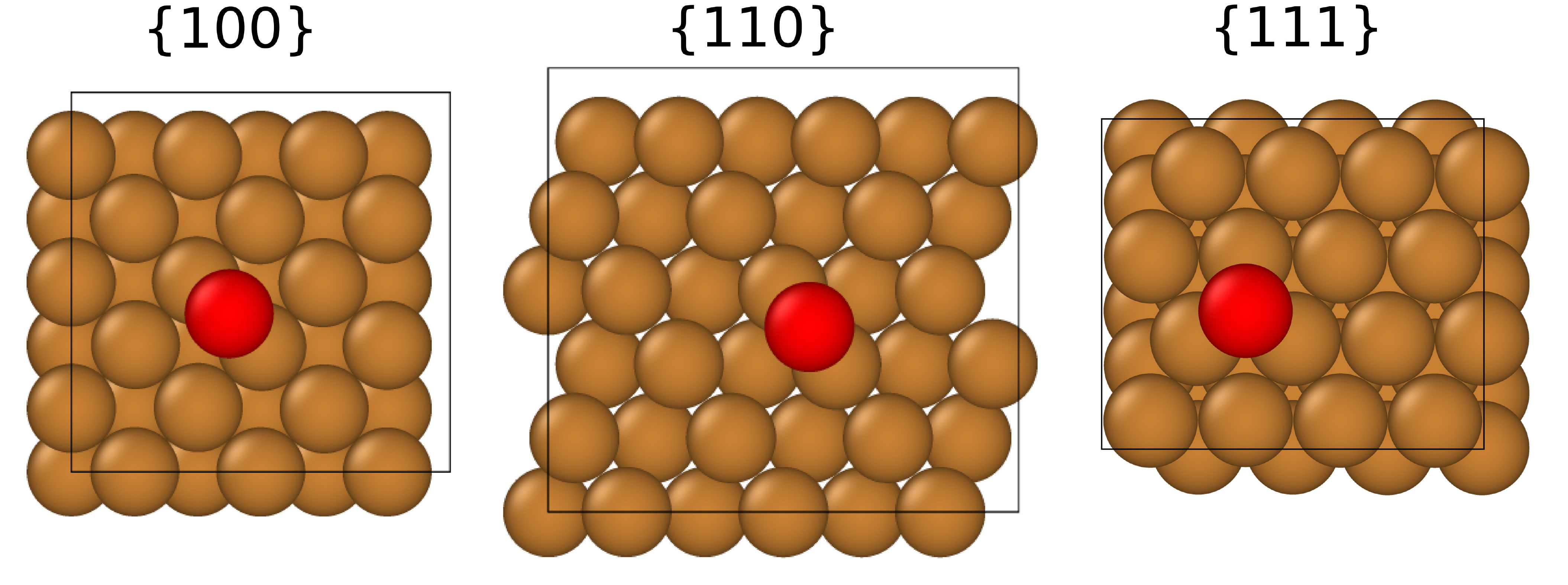}
        \caption{Schematic of the unit cells of the three different surface systems simulated by DFT.}
        \label{fig:DFT_systems}
    \end{figure}
    
    For the calculation of the polarization characteristics of single adatoms placed on Cu surfaces of three different orientations, we used the Vienna Ab Initio Simulation Package~(VASP) with its corresponding pseudopotential data base~\cite{kresse1993ab,kresse1994ab,kresse1996efficient}. We employed the projector-augmented wave~(PAW) potential method~\cite{blochl1994projector,kresse1999ultrasoft} along with the Perdew-Becke-Ernzerhof~(PBE) Generalized Gradient Approximation~(GGA) functional~\cite{perdew1996generalized} to describe the electronic exchange and correlation effects. The cut-off energy was set to 600\,eV. A~$4 \times 4 \times 1$ K-point mesh was used to sample the Brillouin zone according to the Monkhorst-Pack scheme~\cite{monkhorst1976special}. This mesh satisfies $N_k \cdot a \geq 35$\,\AA\ for all directions, where $N_k$ is the number of K-point samples and $a$ is the simulation box size at a given direction. The structures were relaxed until the residual forces were lower than 0.01\,eV/\AA. A vacuum space of at least 25\,\AA\ was appended to the cell and dipole corrections were applied between the slabs. The slab thicknesses were~6,~8, and~8 monoatomic layers for the~\hkl{100},~\hkl{110}, and~\hkl{111} system, respectively.
    
    The polarization characteristics of diffusing adatoms were deduced from the DFT calculations following the methodology developed previously in Ref.~\cite{kyritsakis2019atomistic}. For this, we simulated three different Cu slabs, one for each of the simulated surfaces, as shown in Fig.~\ref{fig:DFT_systems}. An adatom was placed on the top surface in each slab. To obtain the saddle point for the hopping barrier in the~\hkl{100} and~\hkl{110} surfaces, we used the same symmetry considerations as in Ref.~\cite{kyritsakis2019atomistic}, fixing the atom with respect to the lateral directions ($x,\,y$) at the bridge site, while allowing it to relax vertically ($z$). For the~\hkl{111} surface, obtaining the saddle point is much more complex, due to an intermediate local minimum that would occur at the hcp site~\cite{baibuz2018migration}. Since in this work we are interested only in comparison between the polarization characteristics of adatoms deduced from the MD and DFT simulations to an order of magnitude, we can approximate these characteristics on the~\hkl{111} surface with good level of confidence by their values at the lattice site.
    
    Each cell with a specific position of the adatom was calculated for five different values of the electric field between $-2$ and $2$\,GV/m. From these calculations we extracted the systemic dipole moment $\mathcal{M}_s$ ($\mathcal{M}_l$ for the~\hkl{111} surface) and polarizability $\mathcal{A}_s$ ($\mathcal{A}_l$ for the~\hkl{111} surface) by fitting a parabola to the corresponding field-energy curve, as prescribed in Ref.~\cite{kyritsakis2019atomistic}. Additionally, we calculated the reference values $\mathcal{M}_r$ and $\mathcal{A}_r$ for all the surfaces with no adatoms present.
    
  \section{Results}
  \label{sec:results}
  
  \subsection{Accelerated Molecular Dynamics}
    
\begin{figure}
      \centering
      \begin{subfigure}{0.65\linewidth}
        \includegraphics[width=\linewidth]{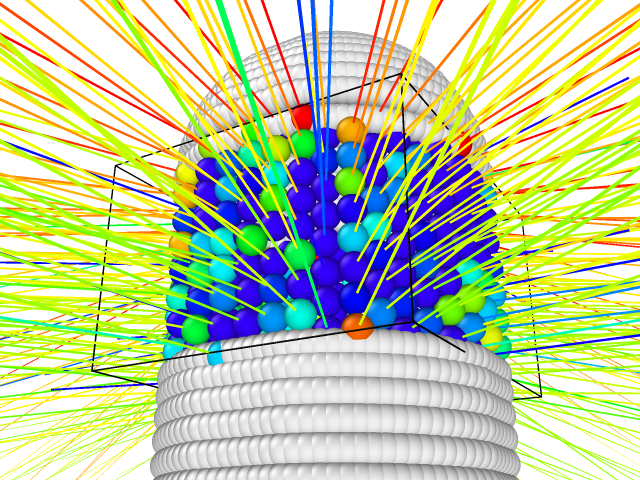}
      \end{subfigure}
      \par\hrulefill\par
      \begin{subfigure}{0.65\linewidth}
        \includegraphics[width=\linewidth]{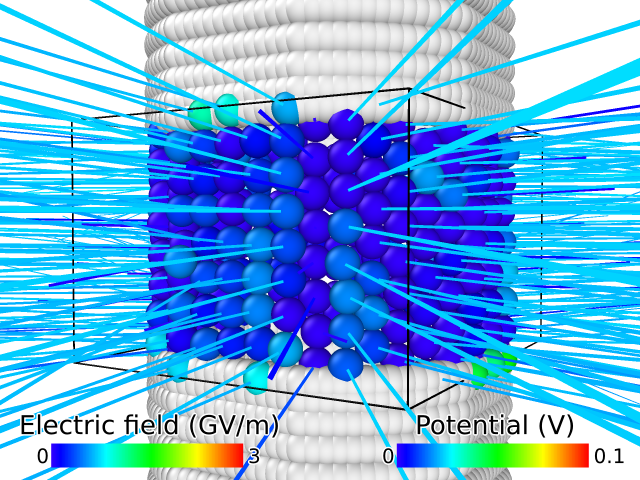}
      \end{subfigure}
      \par\hrulefill\par
      \begin{subfigure}{0.65\linewidth}
        \includegraphics[width=\linewidth]{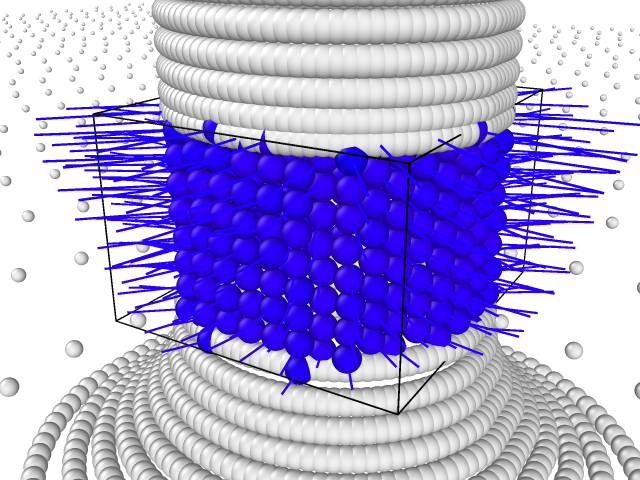}
      \end{subfigure}
      \par\hrulefill
      \par\bigskip
      \begin{subfigure}{0.65\linewidth}
        \includegraphics[width=\linewidth]{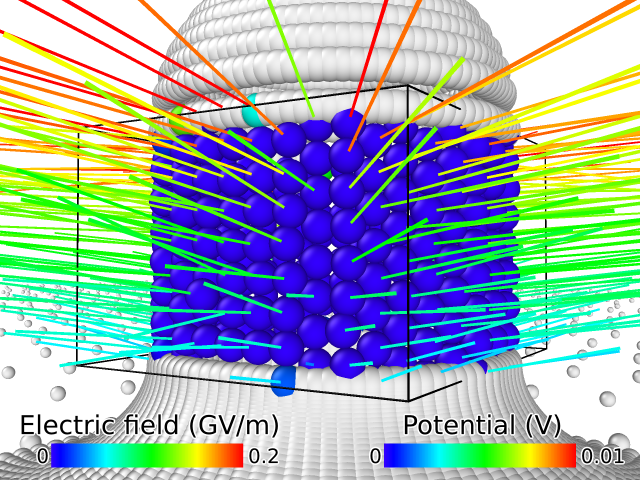}
      \end{subfigure}
      \caption{Electric fields and the electrostatic potential in the~\hkl{100} systems with external electric field $F_\mathrm{ext}=100$\,MV/m. The top three panels are from the tall nanotip (93\,nm) system, and the bottom-most is from the short nanotip (5\,nm). The field vectors point toward the surface atoms, the arrowheads not visible. Atoms are colored according to their potential, and the field vectors are colored according to their magnitude. In the top three panels, the scales of the potential and the field vectors are kept the same, to emphasize how the absolute value of the field also increases in the tall nanotip system. In the bottom-most panel, the magnitudes are scaled up so that the field vectors and the potential coloring are better visible.}
      \label{fig:fieldscharges}
    \end{figure}

    Examples of the fields and the electrostatic potentials of the molecular dynamics~(MD) region in different extended geometries are shown in Fig.~\ref{fig:fieldscharges}. The electric field becomes stronger toward the positive $z$-coordinate, demonstrating the electric field gradient. In the figure, the atoms are colored according to the value of the electrostatic potential and the arrows according to the value of the electric field. The arrows are directed toward the surface, although the arrowheads are not visible, for clarity of the image.
    
    As one can see, the color of the atoms varies, which indicates that the potential on different atoms has different values despite the Dirichlet boundary condition applied at the materials surface. This is due to the potential at each atomic position being calculated as a weighted average over the nodes of the mesh cell where the atom resides. The vacuum nodes contribute non-zero values to the potential, raising its value over the one assumed in the FEM solver. Moreover, there is a tilt in the vectors of the electric field with respect to the normal of the surface, especially evident in the short nanotip system; while the electric field is always perpendicular to the FEM mesh surface, the field at the atomic positions will obtain a parallel component from the vacuum nodes (see~\ref{sec:curl}). These effects are unavoidable at the junction between discrete atom system with atomic resolution and the finite elements of a continuum mesh.
    
    Finally, we remind here that the MD region is wrapped by periodic boundary condition in the $z$-direction, leading to a discontinuity in the electric field when the boundary is crossed. The effect of this discontinuity is expected to be insignificant, as was explained in Sec.~\ref{subsec:fem}.

    Tab.~\ref{tab:factors} summarizes the electric field condition in the nanotips of different geometries. The field enhancement $\beta$ is calculated as
    \begin{equation}
      \label{eq:beta}
      \beta = \frac{F_\mathrm{MD}}{F_\mathrm{ext}}
    \end{equation}
    where $F_\mathrm{MD}$ is the mean magnitude of the electric field that the surface atoms experience in the entire MD region and $F_\mathrm{ext}$ is the magnitude of the applied external electric field. Note that at the bottom of the tall nanotip, the field is in fact suppressed by a factor of~\textapprox0.4, while in the other geometries it is promoted.
    
\begin{figure}
      \centering
      \input{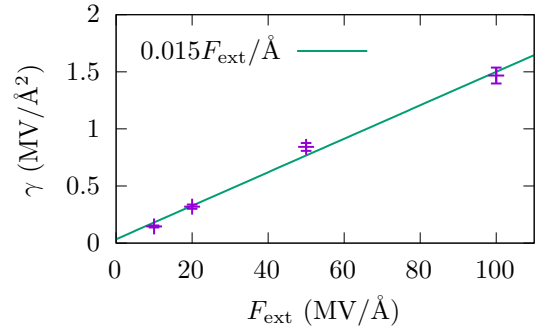}
      \caption{The gradient of the electric field $\gamma$ in the MD system with adatoms on the~\hkl{100} surfaces, placed in the middle of the tall nanotip, as a function of the applied external electric field $F_\mathrm{ext}$ =~1\,V/\AA\ =~10\,GV/m. Magenta points are the simulation data with almost invisible error bars of one standard deviation, and the green line is the linear fit. The slope $s$ of the fitted line is~0.015\,\AA$^{-1}$. All slopes are tabulated in Tab.~\ref{tab:factors}.}
      \label{fig:slope}
    \end{figure}

    We calculate the field gradients as
    \begin{equation}
      \label{eq:gamma}
      \gamma = \frac{F_{r,\,\mathrm{top}}-F_{r,\,\mathrm{bottom}}}{h}
    \end{equation}
    where $F_{r,\,\mathrm{top}}$ is the mean magnitude of the radial component of the electric field in the topmost atomic layer of the MD region, $F_{r,\,\mathrm{bottom}}$ the same in the bottom-most layer, and $h$ the height of the MD region. These gradients are different in different parts of the tip as it depends on the how close to the protrusion tip the layer simulated by MD is. This value approximately equals to $\gamma_{rz}$ of Eq.~\eref{eq:gammatensor}. The gradients are proportional to the applied electric field $F_\mathrm{ext}$ in a given geometry; the tabulated gradients are in fact the slopes $s$ of $\gamma = sF_\mathrm{ext}$ (see figure \ref{fig:slope}).
    \begin{table}
      \centering
      \caption{Field enhancement factors $\beta$ and the radial components of the field gradient slopes $s$ (see Fig.~\ref{fig:slope}) in different geometries. The gradients are expressed in terms of the magnitude of the external electric field $F_\mathrm{ext}$: e.g. the gradient across the~\hkl{100} surface MD region in the middle of the nanotip in~$F_\mathrm{ext}=1$\,V/\AA\ would be~0.015\,V/\AA$^2$.}
      \label{tab:factors}
      \begin{tabular}{llrrr}
        \toprule
        Surface                       & Geometry      & $\beta$ & $s \left(\mathrm{\AA}^{-1}\right)$ \\
        \midrule
        \multirow{4}{*}{\hkl{100}}    & Tall, top     & 18.98   & 0.308    \\
                                      & Tall, middle  &  6.03   & 0.015    \\
                                      & Tall, bottom  &  0.37   & 0.012    \\
                                      & Short, middle &  1.05   & 0.044    \\
        \midrule
        \multirow{4}{*}{\hkl{110}}    & Tall, top     & 21.21   & 0.248    \\
                                      & Tall, middle  &  5.96   & 0.010    \\
                                      & Tall, bottom  &  0.41   & 0.012    \\
                                      & Short, middle &  1.15   & 0.050    \\
        \midrule
        \multirow{4}{*}{\hkl{111}}    & Tall, top     & 19.06   & 0.217    \\
                                      & Tall, middle  &  5.61   & 0.007    \\
                                      & Tall, bottom  &  0.38   & 0.011    \\
                                      & Short, middle &  1.08   & 0.050    \\
        \bottomrule
      \end{tabular}
    \end{table}
    
\begin{figure}
      \centering
      \input{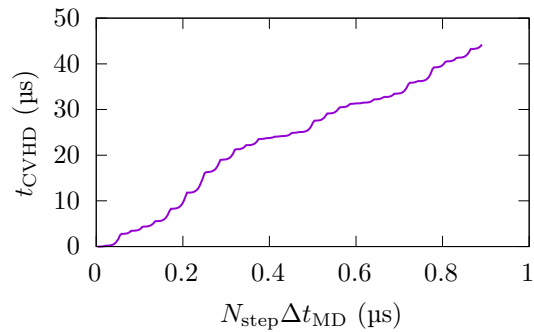}
      \caption{Accelerated ``CVHD time'' as a function of the elapsed ``MD time'', i.e. the number of timesteps $N_\mathrm{step}$ times the MD timestep length $\Delta t_\mathrm{MD}$.}
      \label{fig:acceleration}
    \end{figure}

    Since we aim to analyze the surface atom diffusion bias in terms of drift velocity, it is important to know the exact value of the time elapsed in each simulations. While this information can be obtained directly from MD simulations for the nanotips with~\hkl{110} and~\hkl{111} side facets, the time advance during adatom diffusion on the~\hkl{100} surface requires a special attention as it was discussed in Sec.~\ref{subsec:cvhd}. In Fig.~\ref{fig:acceleration} we show the advantage of the use of the collective variable~-driven hyperdynamics~(CVHD). One can see that for the surface diffusion on the~\hkl{100} surface, the simulation time advances approximately~50--60 times faster than $N_\mathrm{step}\Delta t_\mathrm{MD}$. We use this accelerated time in analysis of the drift velocity of the biased under the electric field diffusion.
    
\begin{figure}
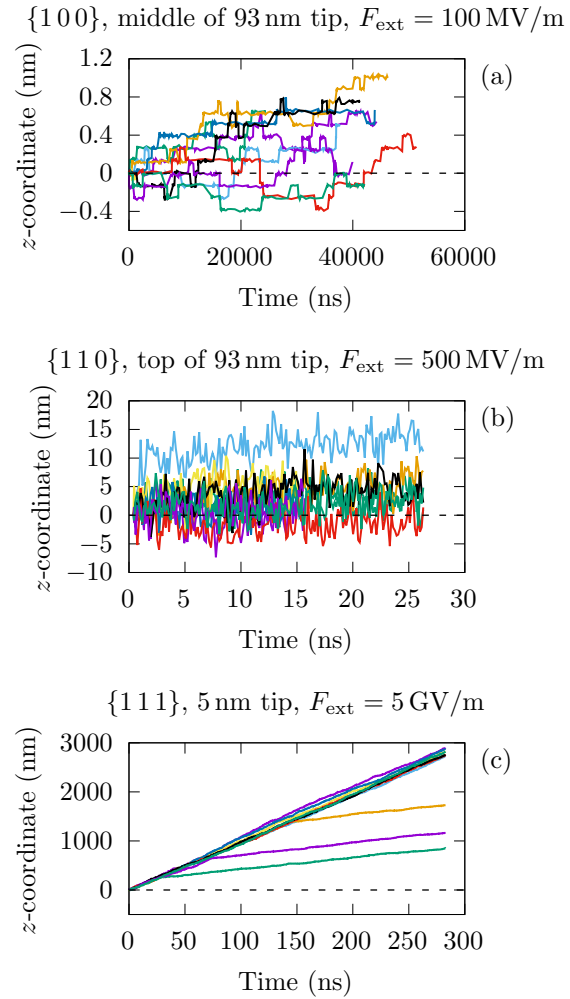

      \begin{subfigure}{\linewidth}
        \centering
        \input{100_z.tex}
      \end{subfigure}
      \begin{subfigure}{\linewidth}
        \centering
        \input{110_z.tex}
      \end{subfigure}
      \begin{subfigure}{\linewidth}
        \centering
        \input{111_z.tex}
      \end{subfigure}
      \caption{Examples of the average $z$-coordinate evolution of the two adatoms diffusing on different surfaces under different electric fields. Panel~(a) shows all~10 runs on the~\hkl{100} surface in the middle of the~93\,nm nanotip system, under an applied field of~100\,MV/m. Panel~(b) shows the same on the~\hkl{110} surface at the top of the nanotip, at~1\,GV/m, and panel~(c) on the~\hkl{111} surface of the~5\,nm nanotip, at~5\,GV/m field. The three lines that diverge from the overall trend in panel~(c) are the simulations where the two adatoms formed a dimer and continued diffusion with a lower bias. Note that the linear increase of the $z$-coordinate in the case of the \hkl{111} surface indicates the strongest bias effect of the electric field gradient on diffusional (i.e. random) hopping of adatoms between the lattice sites on this surface.}
      \label{fig:z}
    \end{figure}

    Examples of the evolution of the average $z$-coordinate of the adatoms on different surfaces are shown in Fig.~\ref{fig:z}. The different jump rates can be clearly seen here: on the~\hkl{100} surface, individual jumps show up in the average $z$-coordinate despite the very long time scale, while on the other two surfaces the jumps seem to blend together. We note that in~\hkl{110} and~\hkl{111} surface simulations, it was necessary to track all atoms from the surface layer and not only the adatoms that were initially placed on these surfaces to study the diffusion behavior. The reason for this is that many transitions on~\hkl{110} and~\hkl{111} surfaces happen via an exchange event: the diffusing adatom occupies a neighboring lattice site that is temporarily freed up due to thermal vibrations. The adatom becomes trapped and turns into a regular surface atom, while the freed up atom continues migration on the surface until the next exchange event takes place. Such exchange events affected the diffusion mainly on the~\hkl{110} surfaces, since on the close-packed~\hkl{111} hopping diffusion is very fast, and thus exchange events only take place when the adatom reaches sharp edges joining the sides separated by the acute angle of the diamond shaped cross-section (see Fig.~\ref{fig:sections}). In some simulations on the~\hkl{111} surface, the two adatoms met forming a dimer, although originally they were placed on the opposite sides of the nanowire. This situation resulted in changing the value of the drift velocity, as can be seen in the changing of the slope for three curves in figure \ref{fig:z}c. These simulations were excluded from the mean drift velocity analysis. Moreover, some of the nanotips lost their integrity during the simulations because of too high electric fields, and were also excluded them from the analysis.

\begin{figure*}
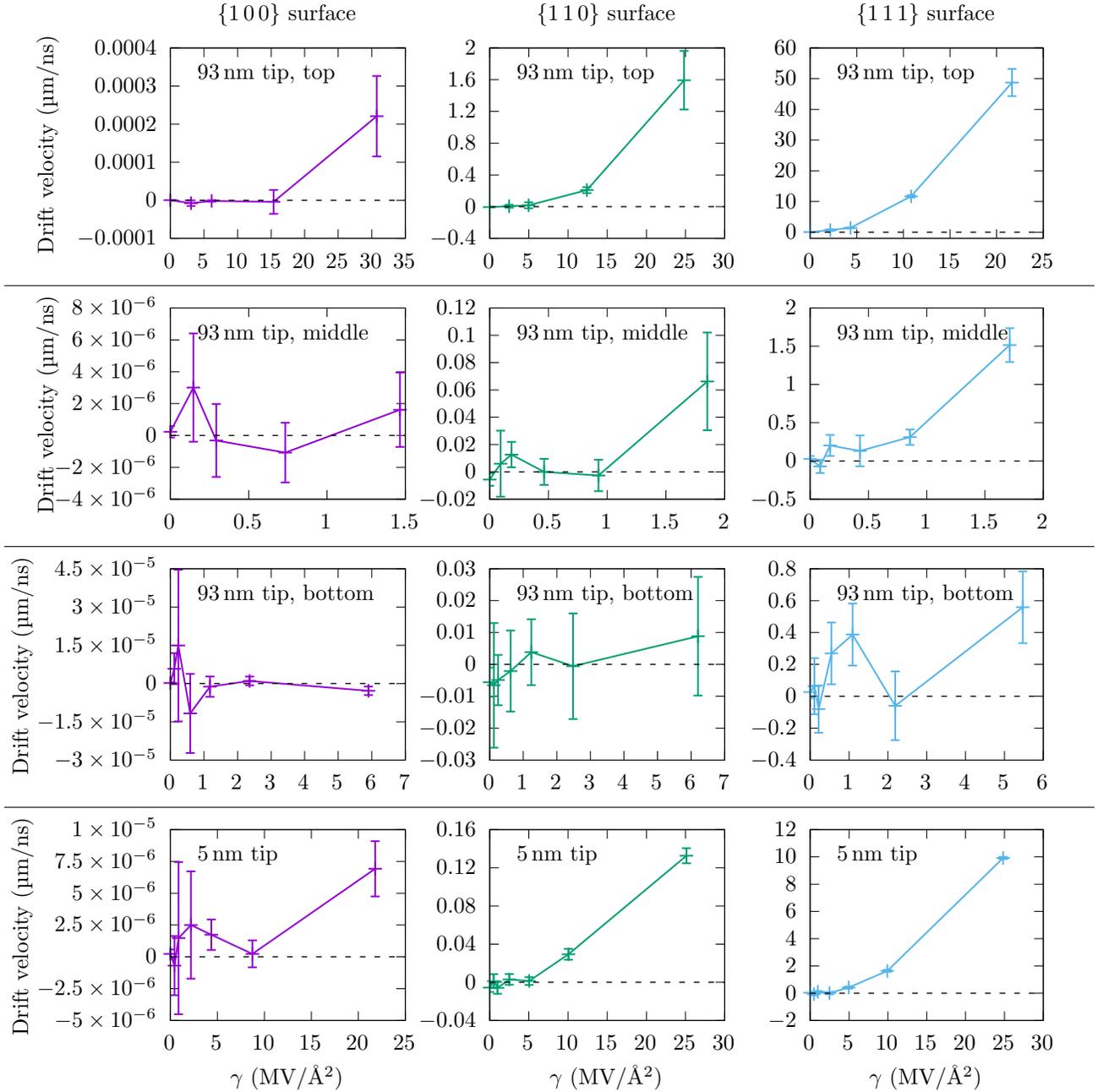

      \centering
      \begin{subfigure}{0.32\textwidth}
        \raggedleft
        \input{100_tall_top_drift_grad.tex}
      \end{subfigure}
      \begin{subfigure}{0.32\textwidth}
        \raggedleft
        \input{110_tall_top_drift_grad.tex}
      \end{subfigure}
      \begin{subfigure}{0.32\textwidth}
        \raggedleft
        \input{111_tall_top_drift_grad.tex}
      \end{subfigure}
      \hrule
      \begin{subfigure}{0.32\textwidth}
        \raggedleft
        \input{100_tall_mid_drift_grad.tex}
      \end{subfigure}
      \begin{subfigure}{0.32\textwidth}
        \raggedleft
        \input{110_tall_mid_drift_grad.tex}
      \end{subfigure}
      \begin{subfigure}{0.32\textwidth}
        \raggedleft
        \input{111_tall_mid_drift_grad.tex}
      \end{subfigure}
      \hrule
      \begin{subfigure}{0.32\textwidth}
        \raggedleft
        \input{100_tall_bottom_drift_grad.tex}
      \end{subfigure}
      \begin{subfigure}{0.32\textwidth}
        \raggedleft
        \input{110_tall_bottom_drift_grad.tex}
      \end{subfigure}
      \begin{subfigure}{0.32\textwidth}
        \raggedleft
        \input{111_tall_bottom_drift_grad.tex}
      \end{subfigure}
      \hrule
      \begin{subfigure}{0.32\textwidth}
        \raggedleft
        \input{100_short_mid_drift_grad.tex}
      \end{subfigure}
      \begin{subfigure}{0.32\textwidth}
        \raggedleft
        \input{110_short_mid_drift_grad.tex}
      \end{subfigure}
      \begin{subfigure}{0.32\textwidth}
        \raggedleft
        \input{111_short_mid_drift_grad.tex}
      \end{subfigure}
      \caption{Average surface diffusion drift velocity in different  geometries and on different surfaces. Error bars are one standard deviation. Note that also the local electric field $F$ varies linearly with the gradient in each plot.}
      \label{fig:drift}
    \end{figure*}

    In Fig.~\ref{fig:drift} we summarize the main results of the current study. Here we show the drift velocity as a function of the electric field gradient $\gamma$ as calculated in Eq.~\eref{eq:gamma}. The drift velocity is calculated as the $z$-displacement of adatoms at the end of the simulation divided by the simulation time. The obtained velocities are averaged over~10 repetitions in each case. The three top rows show results for a tall nanotip with the height of~93 nm, while the diffusion processes were simulated in different parts of the tip: the top, the middle and the bottom from the top row down, respectively. The last row shows the results for a short nanotip of~5 nm in height. The results are also organized in columns: the drift velocity for the surface diffusion on the~\hkl{100},~\hkl{110} and~\hkl{111} surfaces are shown in columns from left to right, respectively.
    
    We observe that on the~\hkl{110} and the~\hkl{111} surfaces (the two rightmost columns), the diffusion bias grows as a function of gradient $\gamma$ in all systems except for the bottom of the 93\,nm nanotip (the second row from the bottom). On the~\hkl{100} surface (the first column), a clear bias can only be seen at high gradient at the top of the 93\,nm nanotip.
    
    The differences within each surface (each column) can be attributed to the different absolute values of the electric field in these systems. As shown in Tab.~\ref{tab:factors}, field enhancement $\beta$ is the highest at the top of the 93\,nm tip, and smallest (less than 1) at the bottom. The differences between surfaces can be explained by different jump rates. See Sec.~\ref{sec:discussions} for further discussions.

  \section{Discussions}
  \label{sec:discussions}

    In our simulations, we observed a clear bias in the migration of adatoms on the three lowest-index (\hkl{100},~\hkl{110}, and~\hkl{111}) Cu surfaces under an electric field gradient. This mechanism promotes sharpening of field-enhancing surface features by adding a bias in the random walk migration of atoms on material surface toward places where the field is higher, such as corners and vertices of protrusions. Sharpening will further strengthen field enhancement and increase gradients, thus creating a positive feedback loop. On the cathode side of the electric system, the sharpening mechanism could promote the growth of small surface roughness into field-emitters with another feedback loop (see for detail Ref.~\cite{kyritsakis2018thermal}) activated: field emission current generates heat that increases the resistivity of the structure, leading to stronger heating and eventually a runaway evaporation of the tip. After evaporation, the biased diffusive process would start to regenerate the sharpness of the remaining protrusion.
    
    The diffusion bias can be explained by the modification of the migration energy barriers by the electric field and its gradient, through the dipole moment and polarizability characteristics of the surface (Eq.~\ref{eq:modbarrier}). The unbiased ($\gamma=0$) jump rate $\Gamma$ is defined by the Arrhenius equation
    \begin{equation}
      \label{eq:arrhenius}
      \Gamma = \nu \exp\left(-\frac{E_\mathrm{m}-\mathcal{M}_\mathrm{sl}F - \frac{\mathcal{A}_\mathrm{sl}}{2}F^2}{k_\mathrm{B}T}\right),
    \end{equation}
    where $\nu$ is assumed to be a field-independent prefactor and $E_\mathrm{m}$ is the migration energy barrier in the absence of electric field. The expected displacement $\left\langle x \right\rangle_\mathrm{b}$  of the adatom after time $\tau$ is
    \begin{equation}
      \label{eq:totalbias}
      \left\langle x \right\rangle_\mathrm{b} = 2\tau l\Gamma\sinh\left(l\gamma\frac{\mathcal{M}_\mathrm{sr} + \mathcal{A}_\mathrm{sr}F}{k_\mathrm{B}T}\right).
    \end{equation}
    It can be seen that the mean displacement is proportional to the unbiased jump rate $\Gamma$. This explains the large differences between the observed drift velocity between surfaces (see Fig.~\ref{fig:drift}): $E_\mathrm{m}$ on the~\hkl{100}, the~\hkl{110}, and the~\hkl{111} surface, given by the MD-MC-CEM potential we used, is~0.52\,eV,~0.25\,eV, and~0.04\,eV, respectively. Without applied electric field, at 300\,K temperature, migration on the~\hkl{110} surface would be approx.~35\,000 times faster, and on the~\hkl{111} surface~$10^8$ times faster than on the~\hkl{100} surface; thus, the bias can be expect to differ multiple orders of magnitude between different surfaces.

    The differences between the bias on the same surface in different geometries are due to the different values of local electric field $F$ in these systems. For instance, the field enhancement factor $\beta$ is 20 times higher at the top of the 93\,nm tip than in the 5\,nm tip, leading to 5--10 times larger drift velocity in the~\hkl{110} and~\hkl{111} systems.
    
    As can be seen from Eq.~\eref{eq:totalbias}, the mean displacement of the adatom depends on the polarization characteristics of the surface, namely the permanent dipole moment difference $\mathcal{M}_\mathrm{sr}\equiv\mathcal{M}_\mathrm{s}-\mathcal{M}_\mathrm{r}$ and  the polarizability difference $\mathcal{A}\equiv\mathcal{A}_\mathrm{s}-\mathcal{A}_\mathrm{r}$. Subscript s stands for the saddle point of the migration event, and r for the flat surface reference system. We can estimate these characteristics from the diffusion bias in MD simulations, and compare them to values calculated directly in DFT. This estimation can be done independently of the migration rate $\Gamma$ and the simulation time, by dividing Eq.~\eref{eq:totalbias} by the mean square displacement $\left\langle x^2 \right\rangle = \tau l^2\Gamma$:
    \begin{equation}
      \label{eq:bias}
      \frac{\langle x\rangle_\mathrm{b}}{\left\langle x^2\right\rangle} =            \frac{2}{l}\sinh\left(l\gamma\frac{\mathcal{M}_\mathrm{sr}+\mathcal{A}_\mathrm{sr}F}{2k_\mathrm{B}T}\right)
    \end{equation}
    We can obtain estimates for $\mathcal{M}_\mathrm{sr}$ and $\mathcal{A}_\mathrm{sr}$ on each surface by fitting Eq.~\eref{eq:bias} to the observed $\langle x\rangle_\mathrm{b}/\langle x^2\rangle$ at each $(F,\,\gamma)$-point. Note that in a given geometry, $F$ and $\gamma$ are both directly proportional to the applied field $F_\mathrm{ext}$, by factors $\beta$ and $s$ of Tab.~\ref{tab:factors}, respectively; thereby, each combination of tip and placement (panel of Fig.~\ref{fig:drift}) only gives a one-dimensional slice in the $(F,\,\gamma)$-space. Thus, to fit the two-dimensional function of Eq.~\eref{eq:bias}, all of the results at given surface must be used in each fit.
    
    The fitting results for $\mathcal{M}_\mathrm{sr}$ and $\mathcal{A}_\mathrm{sr}$ to our data on the three surfaces are shown in Tab.~\ref{tab:msr_asr}, along with the corresponding values calculated directly by DFT according to the methods described in section \ref{DFT}. Unfortunately, we were able to obtain the results for the~\hkl{100} surface only within very large error bars~(\textapprox 100\,\%). This is explained by the limited statistics resulting from the low jump rate on this surface, even when accelerated with CVHD. A further study with higher statistics and longer time spans, and/or stronger electric field gradients would likely permit the fitting of $\mathcal{M}_\mathrm{sr}$ and $\mathcal{A}_\mathrm{sr}$ more accurately. In this study, the agreement of these parameters is qualitative compared to the~\hkl{110} and~\hkl{111} surfaces.
    
    We see that our method underestimates $\mathcal{A}_\mathrm{sr}$ by about an order of magnitude, and predicts $\mathcal{M}_\mathrm{sr}$ generally to have the opposite sign. This reflects the physical nature of the two different quantities. The permanent dipole moments of adatoms on a surface depends on the surface chemistry, and it is therefore impossible to capture them by a purely electrostatic approach. Likewise, the low-field regime diffusion bias toward the weaker field on the cathode side due to the permanent dipole moment, predicted by theory, will not be observed in this model. In high fields, the bias is turned toward the stronger field by the effective adatom polarizability $\mathcal{A}_\mathrm{sr}$, which is directly related to the field-free volume induced by the presence of the adatom, as explained in Ref.~\cite{kyritsakis2019atomistic}. This is a purely electrostatic effect that must appear in simulations that are coupled with electrostatic field calculations. The underestimation of $\mathcal{A}_\mathrm{sr}$ in MD compared to DFT can be partially explained by the different geometry: in MD, we used a cylindrical atomic system with a needle-like FEM extension to maximize the electric field gradient, while the DFT calculation does not need a gradient at all, and thus we used a slab system for computational efficiency. The cylindrical MD system can be expected to polarize differently from a slab. The correct tendency which we observe in Tab.~\ref{tab:msr_asr} confirms that the developed approach indeed is able to reproduce biased effect of surface diffusion in the presence of electrostatic field gradients.
    \begin{table}
      \centering
      \caption{The dipole moment $\mathcal{M}_\mathrm{sr}$ and the polarizability $\mathcal{A}_\mathrm{sr}$ on different surfaces in this study, obtained by fitting Eq.~\eref{eq:bias} to the mean atomic displacement and directly from DFT calculations. 
     The error bars are given by the standard error of the mean. The large error bars for the~\hkl{100} surface are due to limited statistics caused by the lower atomic jump rate.}
      \begin{tabular}{@{}l@{\,}c@{\ }c@{\ }c@{\ }c@{}}
        \toprule
        \multirow{2}{*}{Surface} & \multicolumn{2}{c}{$\mathcal{M}_\mathrm{sr}$ (e\AA)} & \multicolumn{2}{c}{$\mathcal{A}_\mathrm{sr}$ (e\AA$^2$/V)} \\
                                   \cmidrule{2-5}
                                 & MD               &  DFT                           & MD              & DFT                                         \\
        \midrule
        \hkl{100}                & $ .0   \pm .2$   & $.106 \pm .003$                & $.1   \pm .1$   & $.27 \pm .02$                               \\
        \hkl{110}                & $-.008 \pm .005$ & $.094 \pm .006$                & $.034 \pm .008$ & $.30 \pm .04$                               \\
        \hkl{111}*               & $-.016 \pm .006$ & $.162 \pm .003$                & $.02  \pm .01$  & $.23 \pm .02$                               \\
        \bottomrule
      \end{tabular}
      \label{tab:msr_asr}
      \footnotesize{*) The saddle point value for~\hkl{111} surface is approximated by the value at lattice position ($\mathcal{M}_\mathrm{lr},\,\mathcal{A}_\mathrm{lr}$).}
    \end{table}

  \section{Conclusions}
  \label{sec:conclusions}

    We have observed biased self-diffusion on Cu surfaces in the presence of electric field gradients in molecular dynamics simulations coupled to the finite element solver for calculation of electric field distribution at a metal surface. We found that the bias is always toward the stronger value of the electric field, which is consistent with previously reported theoretical predictions. The bias was observed to be stronger on~\hkl{111} and~\hkl{110} surfaces, while diffusion on~\hkl{100} was found to be closer to regular unbiased diffusion since the migration energy barriers for self-diffusion on this surface are much higher than on the other two. The mechanism of biased diffusion can contribute to sharpening and regrowth of field-enhancing nanotips that are proposed to provide the particles necessary for plasma formation in vacuum arc breakdowns. Good agreement between the polarization characteristics deduced from the molecular dynamics simulations with the direct density functional theory calculations confirms the validity of the developed approach.

  \section*{Acknowledgements}

    We would like to thank Ekaterina Baibuz for her contributions to data collection and analysis for the manuscript. J. Kimari was supported by a CERN K-contract. Computing resources were provided by the Finnish IT Center for Science~(CSC) and the Finnish Grid and Cloud Infrastructure (persistent identifier urn:nbn:fi:research-infras-2016072533). Y. Wang, A. Kyritsakis and V. Zadin were funded by the European Union’s Horizon 2020 research and innovation programme under grant agreement No 856705.

  \section*{Data availability statement}

    All data that support the findings of this study are included within the article (and any supplementary files).

  \section*{ORCID iDs}

    J. Kimari \orcidlink{0000-0002-4363-1326} \url{https://orcid.org/0000-0002-4363-1326}\\
    A. Kyritsakis \orcidlink{0000-0002-4334-5450} \url{https://orcid.org/0000-0002-4334-5450}\\
    V. Zadin \orcidlink{0000-0003-0590-2583} \url{https://orcid.org/0000-0003-0590-2583}\\
    F. Djurabekova \orcidlink{0000-0002-5828-200X} \url{https://orcid.org/0000-0002-5828-200X}\\

  \appendix

  \section{Electric field outside an equipotential surface}
  \label{sec:curl}

    By definition, the direction of the electric field is always perpendicular to the equipotential surface. This is true in nature as well as in simulations that use the Dirichlet boundary condition.

\begin{figure}
      \centering
      \includegraphics[width=0.6\linewidth]{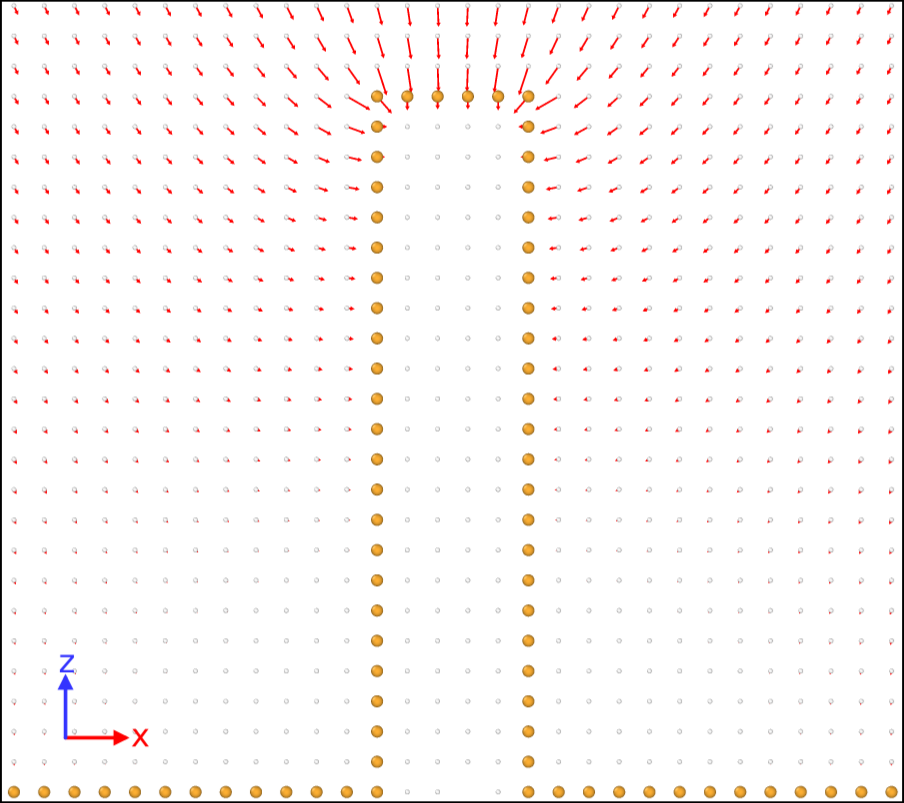}
      \caption{Nanotip on a 2-dimensional conductive (equipotential) surface. The copper-colored atoms mark the location of the surface, and the smaller, white points are additional mesh points for the field calculation. Red arrows show the direction and the relative magnitude of the field.}
      \label{fig:tip}
    \end{figure}

    In the presence of an electric field gradient on the surface, what is the direction of the electric field that the surface \emph{atoms} feel? It is instructive to look at the curl of the electric field $\nabla\times\mathbf{F}$, that has to be zero everywhere in the absence of a changing magnetic field $\mathbf{B}$ by the Maxwell-Faraday equation:
    \begin{equation}
      \label{eq:curl}
      \nabla\times\mathbf{F} = -\frac{\partial\mathbf{B}}{\partial t} = \mathbf{0}
    \end{equation}
    For simplicity, the curl in two dimensions is equal to
    \begin{equation}
      \label{eq:curl2d}
      \nabla\times\mathbf{F} = \left( \frac{\partial F_x}{\partial z} - \frac{\partial F_z}{\partial x}\right) \mathbf{\hat y}
    \end{equation}
    We orient the surface such that it is parallel to the $z$-coordinate and set the electric field to increase toward the positive $z$-coordinate, like on the vertical sides of the system shown in Fig.~\ref{fig:tip}.

\begin{figure}
      \centering
      \includegraphics[width=0.6\linewidth]{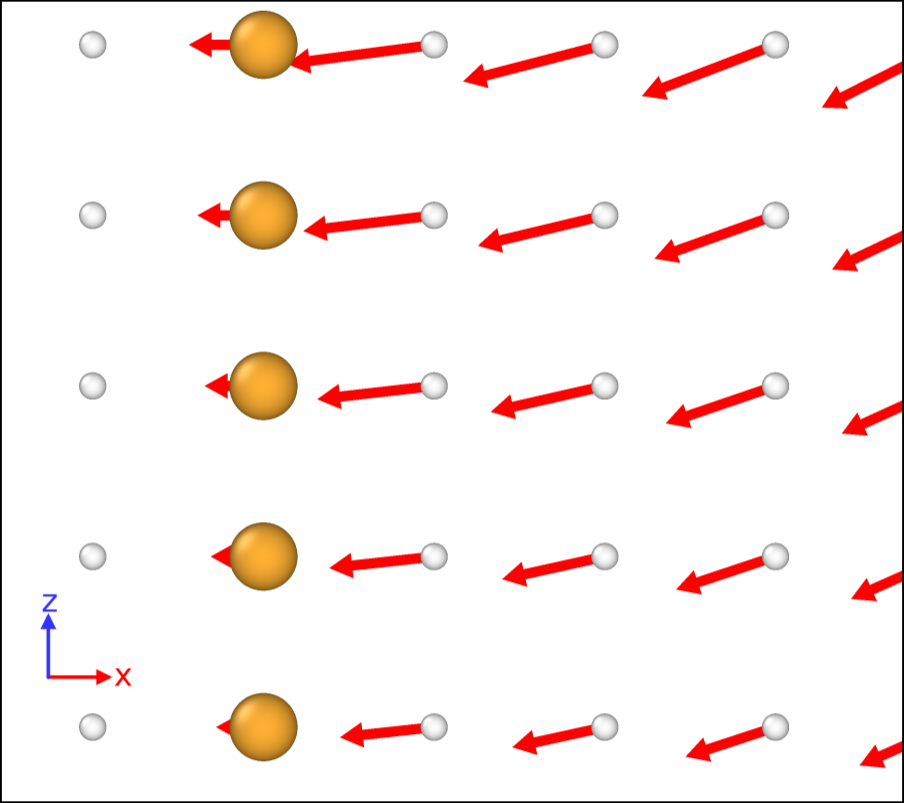}
      \caption{Zoom-in to the surface of the nanotip. Coloring as in Fig.~\ref{fig:tip}. At the first points outside the equipotential surface, the electric field is already tilted downward. This is not an artifact of finite grid spacing, but a physical necessity arising from Eq.~\eref{eq:curl}.}
      \label{fig:curlzoom}
    \end{figure}

    On the righthand side surface of the tip, the term $\sfrac{\partial F_x}{\partial z} < 0$, as the $x$-component is becomes smaller (more negative) when moving toward the positive $z$-coordinate. For Eq.~\eref{eq:curl} to hold, the following must be true:
    \begin{equation}
      \label{eq:curlcomponents}
      \frac{\partial F_z}{\partial x} = \frac{\partial F_x}{\partial z} < 0
    \end{equation}
    In other words, crossing the surface from inside to outside, the $z$-component of the electric field must \emph{decrease}. Since the electric field inside the surface is zero, $F_z$ must be negative already at an infinitesimal distance outside the surface. The same is observed in a numerical finite differences method solution of the 2-dimensional Laplace equation (Eq.~\eref{eq:laplace}) in Fig.~\ref{fig:curlzoom}.

    As the surface atoms occupy a finite volume that extends some distance to the vacuum, the direction of the electric field they feel cannot be perfectly perpendicular to the surface wherever there exists an electric field gradient. The tilt of the field is opposite to the direction of the gradient.
  
\section*{References}

\bibliography{bibliography.bib}

\end{document}